\documentclass[preprint,12pt]{elsarticle}

\usepackage{graphicx}
\usepackage{amssymb}
\usepackage{tikz}

\usepackage{lineno}

\journal{Journal Name}

\begin{document}

\begin{frontmatter}


\title{Projector quantum Monte Carlo with averaged vs explicit spin-orbit effects: applications to tungsten molecular systems}



\author{Cody A. Melton, M. Chandler Bennett, and Lubos Mitas}

\address{North Carolina State University\\
Raleigh, NC 27695-8202}

\begin{abstract}
We present a recently developed projector quantum Monte Carlo method for calculations of electronic structure in systems with spin-orbit interactions. 
The method solves for many-body eigenstates in the presence of 
spin-orbit using the fixed-phase approximation.
The trial wave function is built from two-component spinors and explicit Jastrow correlation factors while the core electrons are eliminated by relativistic effective core potentials with explicit spin-orbit terms. We apply this method to WO and W$_2$ molecules that enable us to build multi-reference wave functions and analyze in detail the impact of both electron correlations and of spin-orbit terms. These developments open new opportunities for calculations of systems with significant spin-orbit effects by
many-body quantum Monte Carlo methods.
\end{abstract}


\end{frontmatter}

\section{Introduction}
\label{sec:intro}
Over the decades, there has been  considerable interest in transition metal
systems due to their intriguing physical properties and their potential applications.
As a results of the versatility of $d$-bonding, a wide variety of physical phenomena have been observed in transition metal compounds such as superconductivity, ferroelectricity, a plethora of magnetic phases and more. 
Molecular nanostructures, surfaces and interfaces offer still more potential uses in both research and applications.
For example, the versatility of the $d$-bonding finds utility in catalysts and bioenzymes \cite{catalyst1,catalyst2}.
For 2D transition metal dichalcogenides, the optical and properties and high electron mobility make them ideal candidates for photonics, optoelectronics, spintronics, etc \cite{tmdc}.
Apart from applications, transition metal compounds are of great theoretical interest due to their strong electronic correlations which are difficult for standard theoretical techniques to describe (e.g. Mott insulators).
Between the family of electronic structure quantum Monte Carlo (QMC) methods, 
the diffusion Monte Carlo (DMC) has had significant success at accurately describing systems with strong correlations, including transition metal compounds \cite{bennett1,feo, cuprate,tm_review,vo2,tio2}.
Despite its successes, DMC traditionally works with many-body Hamiltonians without spin terms.
This implies that important physics of spin-orbit compounds, non-collinear spin phases, etc, has not been accessible to this unique methodology. 
While the strength of the spin-orbit may be rather modest in the 3rd row transition metals, as one begins to consider heavier elements in the 5th row and beyond, scalar relativistic and spin-orbit effects are no longer negligible and can be comparable to that of the correlation effects.
Thus so far, heavy element systems with strong spin-orbit interactions have been left out from electronic structure QMC studies.

Recently, we have generalized DMC algorithm to work with spinorbitals and spinors within a complex wave function framework under the fixed-phase approximation and we named the method fixed-phase spin-orbital diffusion Monte Carlo (FPSODMC) \cite{melton1,melton2}.
Using this construction, we showed that DMC is now capable of treating the spin-orbit terms in the Hamiltonian directly without relying on perturbation theory as is typically done in many mean-field techniques.
We applied the methodology to some simple atoms and molecules, illustrating that DMC is able to directly include the spin-orbit corrections while still accurately describing the electronic correlations.
In this paper, we apply the FNDMC and FPSODMC methods to molecular systems containing tungsten with scalar relativity only (with averaged spin-orbit) 
and  with explicit spin-orbit terms included, respectively.
In particular, we study the tungsten oxide (WO) and tungsten dimer (W$_2$) molecules where limited experimental data exists.

The paper is organized as follows.
First, we briefly introduce the fixed-phase DMC method that must be employed for complex trial wave functions.
After introducing the methodology used for complex wave functions, we introduce a complex representation for the electrons spin degree of freedom which allows for efficient sampling of this space. 
We also describe how the relativistic effects enter directly into the Hamiltonian through the use of non-local relativistic effective core potentials that are used in relativistic quantum chemistry. 
The non-local effective core potentials gives us the freedom to choose between including only scalar relativistic effects with averaged spin-orbit (AREP) or whether to include spin-orbit explicitly and directly
(REP).
We then present our results for the tungsten molecules WO and W$_2$ and compare our results against other theoretical investigations and experiment where applicable.
We then conclude with a summary as well as an outlook on how FPSODMC can be improved for future studies of correlated materials with spin-orbit interactions included.

\section{Fixed-Phase Diffusion Monte Carlo}
\label{sec:fpdmc}
Imaginary time projector methods, like DMC, work to project out the ground state in the infinite limit, namely $\lim\limits_{\tau \rightarrow \infty}e^{-\tau H}\Psi$ $\propto$ $\Psi_0$.
However, in order to avoid the notorious fermion sign problem, one must invoke an approximation utilizing a trial wave function.
Traditionally, this approximation comes in the form of the fixed-node approximation, where the ground state wave function that is projected has the same zeros (nodes) as the trial wave function.
While the unrestricted DMC projection is formally exact, the boundary conditions enforced by the fixed nodes introduce a bias.
This manifests in higher total energy since the fixed-node approximation is variational.
If the trial wave function happens to have the exact nodal structure, the energy obtained is the exact ground state energy.
While fixed-node diffusion Monte Carlo (FNDMC) is the most common implementation \cite{qmcrev}, it cannot be used for inherently complex wave functions as are needed, for example, for spin-orbit applications. 
For complex wave functions, a more general 
elimination of sign/complex amplitudes called the fixed-phase approximation \cite{ortiz} was introduced some time ago.
As outlined below, we use the fixed-phase framework to formulate DMC with variable fermionic spins.

For simplicity, we sketch the fixed-phase
method first considering just the spatial degrees of freedom while the spins will be added later. Let us denote our configuration space as $\mathbf{R} = \{ \mathbf{r}_1,\mathbf{r}_2,\ldots,\mathbf{r}_N \}$. 
The imaginary time Schr\"{o}dinger equation is written as 
\begin{equation}
    -\frac{\partial \Psi(\mathbf{R},\tau)}{\partial \tau } = \left[-\frac{1}{2}\nabla^2_\mathbf{R} + V(\mathbf{R}) + W \right]\Psi(\mathbf{R},\tau)
    \label{eqn:schrodinger}
\end{equation}
where $-1/2\nabla_\mathbf{R}^2$ is the kinetic energy, $V(\mathbf{R})$ is the local potential, and $W$ is any non-local potential such as a pseudopotential (or effective core potential).
Here, we assume a complex wave function $\Psi(\mathbf{R},\tau) = \rho(\mathbf{R},\tau) \exp\left[ i \Phi(\mathbf{R},\tau) \right]$, where $\rho\geq 0$ and $\Phi$ are the amplitude and phase, respectively.
Substituting this into equation (\ref{eqn:schrodinger}) yields two coupled equations for the real and imaginary parts of the Schr\"{o}dinger equation, namely
\begin{eqnarray}
    \label{eqn:real_schrodinger}
    -\frac{\partial \rho}{\partial \tau} &=& -\frac{1}{2}\nabla^2_{\mathbf{R}}\rho + \frac{1}{2}\left|\nabla_{\mathbf{R}} \Phi\right|^2\rho + V(\mathbf{R})\rho + \rm{Re}\left[ \Psi^{-1}W \Psi \right]\rho \\
    \label{eqn:imag_schrodinger}
    -\frac{\partial \Phi}{\partial \tau} &=& -\frac{1}{2}\nabla^2_{\mathbf{R}} \Phi - \rho^{-1}\nabla_{\mathbf{R}}\rho \cdot \nabla_{\mathbf{R}} \Phi + \rm{Im}\left[ \Psi^{-1}W \Psi \right]
\end{eqnarray}
We first note that the expectation value of the Hamiltonian must be real since the Hamiltonian is Hermitian, i.e., the eigenvalue is determined from the real part given by equation (\ref{eqn:real_schrodinger}).
The two coupled equations cannot easily be solved, so we will approximate the phase by a time independent trial wave function $\Psi_T (\mathbf{R}) = \rho_T(\mathbf{R}) \exp\left[ i \Phi_T(\mathbf{R}) \right]$ and only consider solving the real part.
An approximate phase can only raise the energy, so this approximation is variational as given elsewhere \cite{ortiz, melton2}.
An additional approximation must be invoked since we do not know the action of the nonlocal potential $W$ on the wave function $\Psi$ since we are trying to solve for $\Psi$. 
We therefore invoke the locality approximation \cite{locality} and replace $\Psi^{-1}W\Psi$ by $\Psi_T^{-1}W \Psi_T$. 
It can be shown that the error from this approximation goes as $\left( \Psi_T - \Psi_0 \right)^2$ where $\Psi_0$ is the ground state wave function. 
While the localization approximation is not variational, one can restore the upper bound property made by enforcing the so-called T-moves \cite{casula,melton2}.

As is traditionally done in DMC, we perform an importance sampling transformation \cite{reynolds} by multiplying the  real part by the trial amplitude $\rho_T$. 
Coupled with our various approximations, the importance sampled Schr\"{o}dinger equation becomes
\begin{equation}
    -\frac{\partial f}{\partial \tau} = -\frac{1}{2}\nabla^2_{\mathbf{R}}f + \nabla_{R}\left[ f \nabla_{R}\ln \rho_T \right] + E_L(\mathbf{R}) f
    \label{eqn:importance_sampled}
\end{equation}
where the mixed distribution $f$ is given as a product, $f(\mathbf{R},\tau) = \rho_T(\mathbf{R}) \rho(\mathbf{R},\tau)$, and the local energy $E_L(\mathbf{R})$ is 
\begin{equation}
    E_L(\mathbf{R}) = -\frac{1}{2} \frac{\nabla_{\mathbf{R}}^2 \rho_T}{\rho_T} + \frac{1}{2}\left| \nabla_{\mathbf{R}}\Phi_T\right|^2 +  V(\mathbf{R})  + \rm{Re}\left[ \Psi_T^{-1} W \Psi_T \right]
    \label{eqn:local_energy}
\end{equation}
We solve equation (\ref{eqn:importance_sampled}) in integral form
\begin{equation}
    f(\mathbf{R}',\tau +\delta \tau) = \int \textrm{d}\mathbf{R}\; G(\mathbf{R}',\mathbf{R};\delta\tau) f(\mathbf{R},\tau)
    \label{eqn:f_integral}
\end{equation}
where the Green's function can be approximated for small time steps $\delta \tau$ as
\begin{eqnarray}
    G(\mathbf{R}',\mathbf{R};\delta \tau) \approx (2\pi\delta\tau)^{-3N/2}\exp\left[  \frac{-\left|\mathbf{R}'-\mathbf{R}-\delta\tau \nabla_\mathbf{R}\ln \rho_T \right|^2}{2\delta\tau} \right] \nonumber\\
    \times \exp\left[ -\frac{\delta \tau}{2}\left(E_L(\mathbf{R}')+E_L(\mathbf{R}) - 2E_T\right) \right]
    \label{eqn:greens_fn}
\end{eqnarray}
where we have introduced an energy offset $E_T$ to the Hamiltonian and $N$ is the number of electrons.
The error associated with this approximate Green's function goes as $O(\delta \tau^3)$ and can easily be extrapolated to zero.
By repeated application of the Green's function, we can step forward in imaginary time to achieve the steady state solution which corresponds to the ground state wave function and energy.

\section{Spin-Orbit Interactions and Dynamic Spins}
\label{sec:so_dyn_spin}
In the previous discussion, we have restricted ourselves to static spins and only reference the spatial configuration space.
In our previous works \cite{melton1,melton2,melton3,melton4}, significant descriptions for the  introduction of dynamic spins into the configuration space has been given, so we only briefly describe it here.

If we consider an arbitrary one-particle spinor
\begin{equation}
    \psi(\mathbf{r},s) = \alpha \phi^\uparrow(\mathbf{r})\chi^\uparrow(s) + \beta \phi^\downarrow(\mathbf{r})\chi^\downarrow(s)
    \label{eqn:spinor}
\end{equation}
where $\alpha,\beta$ are complex constants, $\phi^{\uparrow/\downarrow}$ are (complex) spatial orbitals and $\chi^{\uparrow/\downarrow}$ are the spin functions, where $\chi^{\uparrow/\downarrow}(s=\pm1/2)$ is the standard $S_z$ representation for the spin degrees of freedom.
DMC, however, traditionally works in a configuration space that is continuous whereas the spin variables are discrete.
This begs the question as to whether the spin variables can be made continuous in order to sample them in a similar fashion to the spatial degrees of freedom.
Indeed this can be achieved, through the introduction a complex representation for the spin states as well as a spin ``kinetic energy.''
The new spin representation is given as 
\begin{equation}
    \chi^\uparrow(s) = e^{i s}, \; \chi^\downarrow(s) = e^{-is}, \;\;\;\;\; s\in [0,2\pi)
    \label{eqn:spin_rep}
\end{equation}
which preserves the orthogonality of the $\chi^{\uparrow/\downarrow}$ states.
It should be noted that using the discrete representation, the determinants in the trial wave function in Eq. \ref{eqn:trial_func} breaks up into a product of spin up and spin down determinants \cite{melton4}, whereas using our complex spin representation we retain a single determinant.

Clearly, the results obtained depend on the quality of the trial phase, and thus, the trial wave function. 
We write the many-body trial wave function in the multi-reference Slater-Jastrow form
\begin{equation}
    \Psi_T(\mathbf{R,S}) = \exp\left[ J(\mathbf{R} )\right] \sum_{k=1}^M c_k\rm{det}_k\left[ \psi_i(\mathbf{r}_j,s_j)\right]
    \label{eqn:trial_func}
\end{equation}
where $M$ is the number of determinants, $\psi_i(\mathbf{r}_j,s_j)$ is the $i$th single-particle spinor for electron $j$, and $J(\mathbf{R})$ is the Jastrow factor which explicitly includes inter-particle correlations to the wave function.
If the trial phase happens to be the exact phase of a given symmetry, then the algorithm is exact and will reproduce the exact ground state energy within that symmetry channel.

In order to sample the continuous spins, we add  to the Hamiltonian a term 
\begin{equation}
    T_\mathbf{S} = \sum_{i=1}^{N} T_{s_i}, \;\;\;\;\; T_{s_i} = -\frac{1}{2\mu_s}\left[ \frac{\partial^2}{\partial s_i^2} + 1 \right]
    \label{eqn:spin_kin}
\end{equation}
that generates the spin evolution while
$\mu_s$ is a ``spin mass''. Since
$T_{s_i} \psi(\mathbf{r}_i,s_i) = 0$, 
it also annihilates any trial wave function generally written as a linear combination of spinor determinants. 
Note that the spin mass $\mu_s$ results in an effective time step for the spin degrees of freedom, namely $\delta\tau_s = \delta\tau/\mu_s$. 
With the addition of this spin kinetic energy, we are simulating a Hamiltonian of the form $H' = H + T_{\mathbf{S}}(\mu_s)$.
As we discussed elsewhere \cite{melton2, melton4} the ratio of spin time step and the spatial time step can vary and provides different limits that imply faster (or slower) spin evolution with regard to the spatial degrees of freedom. 
Here we employ
the limit of $H' \rightarrow H$ by taking $\mu_s \rightarrow \infty$, ie, $\tau_s \to 0$, which corresponds to slowly evolving spins compared to the spatial degrees of freedom and enables to recover the 
original Hamiltonian without the contamination from the spin kinetic energy.

The spin kinetic energy acts to add dynamics for the spin degrees of freedom, but thus far no reference to the spin-orbit interaction has been made. 
For matter at the nuclei and electrons resolution a straightforward way to introduce scalar and spin-orbit effects is through the use of relativistic effective core potentials (REPs) \cite{rel_pp, dolgrev}. 
The motivation for this is twofold. 
First, all-electron calculations within DMC scales poorly with the atomic number, namely $Z^{5.5-6.5}$ whereas they scale as $N^{3}_{val}$ for the valence electrons when an ECP is utilized \cite{ceperley86,hammond87}. Since the core 
electrons do not affect the valence properties directly
 it is favorable to remove the cores by an accurate effective model.
This allows DMC to treat larger and more complicated systems with much more favorable scaling and resolution 
for valence energy differences. 
Second, if one considers relativity, one needs to solve the all-electron Hamiltonian within the four-component spinor formalism. 
For the deepest core electrons, all four components of the spinors can be non-zero.
For valence electrons that are relevant in bonding, the 4-component spinors essentially decouple into major and minor 2-component spinors. 
The minor component spinors can be eliminated by using several approximations/transformations of the relativistic Hamiltonian (such as  DKH or X2C), leaving only 2-component spinors for the valence electrons. 
If the core electrons are projected out and replaced with an effective operator, what is left are 2-component spinors for the valence electrons moving in an effective potential that represents the impact of the relativistic core on degrees of freedom. 
Given our two-component spinor representation above, we simply include these effective core operators into our Hamiltonian and drop the 
atomic cores and corresponding degrees of freedom.

For any Hamiltonian that does not depend on spin, both non-relativistic and scalar relativistic with averaged spin-orbit (AREP), the nonlocal ECP typically is written as a projector with spherical harmonics with a local radial dependence
so that both spins feel the same projector.
In cases where spin is included in the Hamiltonian such as spin-orbit, spin no longer commutes with the Hamiltonian and only the total angular momentum ${\bf J} ={\bf L}+{\bf S}$ is a good quantum number. 
Thus, the REPs utilize a more general form of the projectors,
\begin{equation}
W^{REP}_i = \sum_\ell \sum_{j=|\ell-1/2|}^{|\ell+1/2|}\sum_{m_j = -j}^{j} W_{\ell j}^{REP}(r_i) | \ell j m_j \rangle \langle \ell j m_j |
    \label{eqn:REP}
\end{equation}
where $|\ell j m_j \rangle$ are the so-called spin spherical harmonics and the radial dependence is typically parametrized by
\begin{equation}
    W_{\ell j}^{REP}(r_i) = \frac{1}{r_{i}^2} \sum_{\alpha} A_{\ell j \alpha} r_{i}^{n_{\ell j \alpha}}  \exp\left[ -B_{\ell j \alpha} r_i^2 \right]
    \label{eqn:radial_gaussian}
\end{equation}
This form of the RECP captures both scalar and spin-orbit relativistic effects, and parameterizations have been generated by several research groups \cite{stu,crenbl}.
The contribution from both scalar and spin-orbit relativistic effects enter into the local energy in equation (\ref{eqn:local_energy})
\begin{eqnarray}
    \textrm{Re}[\frac{W^{REP}\Psi_T}{\Psi_T}] = \textrm{Re}\Big[ \sum_{iI} \sum_{\ell j}W^{REP}_{\ell j}(r_{iI}) \int d\Omega'_{iI}\int ds'  \nonumber \\
    \sum_m \langle \Omega_{iI}s_s | \ell j m\rangle\langle \ell j m | \Omega_{iI}' s_i'\rangle \frac{\Psi_T(\ldots,\mathbf{r}_i',s_i',\ldots)}{\Psi_T(\ldots,\mathbf{r}_i,s_i,\ldots)}\Big]
\end{eqnarray}
where $\mathbf{r} = (r,\Omega)$, $\mathbf{r}' = (r,\Omega')$ and $\Omega$ is the solid angle around a particular nucleus.
While the form above is completely general for an arbitrary spin representation, utilization of the complex spin functions described above yields a simple complex functional form for the spin spherical harmonics \cite{melton2}. 

The spin degrees of freedom then enter the Green's function in a similar way as spatial coordinates. Therefore they appear both in
evolution represented by the diffucion and 
drifts as well as in the local energy.
The evolution of the walker coordinates
(space and spin)
is governed by the Eq. (6).

\section{Results}
\label{sec:results}
We illustrate the method on heavy transition metal molecules, namely tungsten oxide WO and the tungsten dimer W$_2$. 
This is partially driven by the fact that the spinors (orbitals) and corresponding wave functions are much better developed in molecular codes that enable us to study the combined correlation and spin-orbit effects more systematically.
For both cases, we study the molecules using scalar relativistic spin-averaged Hamiltonians in one-component conventional framework as well as with spin-orbit included in two-component framework outlined above.
For scalar relativistic Hamiltonians we use FNDMC whereas for spin-orbit Hamiltonians we utilize FPSODMC \cite{melton1,melton2} using a modified version of \textsc{Qwalk} \cite{qwalk}.
The trial wave functions for the scalar relativistic Hamiltonians use either Hartree-Fock (HF) or Density Functional Theory (DFT) orbitals from \textsc{Gamess} \cite{gamess} to define the nodal surface.
For the FPSODMC, the trial wave functions are generated by the \textsc{Dirac} code \cite{dirac}, and are typically multi-reference in nature. 
For multi-reference wave functions, we use either complete open-shell configuration interaction (COSCI) or larger configuration interaction wave functions that use Dirac HF (DHF) single particle spinors.
In closed-shell systems, the trial wave function is a single determinant and one can use either DHF spinors or relativistic DFT spinors.
For the W atom, we use the REPs from the Stuttgart group \cite{stu} whereas we use the O ECP of Burkatzki, Filippi, and Dolg \cite{bfd}.
\subsection{Tungsten Oxide}
The tungsten atom and many tungsten containing compounds have a different electronic structure than their isovalent counterparts containing Cr and Mo. 
For example, the ground state configuration of W is [Xe$4f^{14}$]$6s^25d^4$, whereas both Cr and Mo have $ns^1(n-1)d^5$ ground state configurations \cite{melton1}.
The favored configuration and state results from subtle balance between electron correlation as well as the splitting due to spin-orbit. Interestingly,
this trend manifests itself also in the
WO molecule.
CrO and MoO have $^5\Pi$ molecular ground states, and some of the original theoretical work on WO predicted a $^5\Pi$ ground state \cite{nelin_bauschlicher}.
However, experimental determination as well as contracted multi-reference configuration interaction (CMRCI) calculations show a $^3\Sigma^-$ ground state for WO \cite{ram2001}.
Our results described below confirm this conclusion using both scalar relativistic and spin-orbit Hamiltonians, indicating that this effect is predominantly driven by correlations with smaller impact from the spin-orbit effects.

We first discuss the determination of the ground state without the spin-orbit interaction. For experiments the estimation 
of the equilibrium bond length
 was found to be 1.65807(6)~\AA\,  \cite{ram2001}.
The CMRCI calculations predict a minimum at 1.67~\AA\,  for the $^3\Sigma$ state, which is only in error by $\sim 0.01$~\AA.
In order to determine the ground state, we calculate the $^3\Sigma$ and $^5\Pi$ states at $r_e = 1.67$~\AA.
We also tested various nodal surfaces, namely PBE and hybrid PBE0 which includes (HF) exchange, which in many cases can have a significant impact on the nodal surface, especially for transition metals.
The results are shown in Table \ref{tab:WO_arep_gs}.
\begin{table}
    \centering
    \caption{Total energies for WO for the $^3\Sigma$ and $^5 \Pi$ states at $r_e = 1.67$~\AA. These results only utilize a scalar relativistic Hamiltonian and neglects spin-orbit.}
    \begin{tabular}{c|ccc}
	\hline\hline
	$\Psi_T$        & SCF & VMC & FN-DMC \\
	\hline
	$^3\Sigma$ PBE  & -83.501631 & -83.2377(2) & -83.3149(3) \\
	$^3\Sigma$ PBE0 & -83.429649 & -83.2397(1) & -83.3130(3) \\
	$^5\Pi$ PBE     & -83.484707 & -83.2227(2) & -83.2870(4) \\
	$^5\Pi$ PBE0    & -83.398745 & -83.2244(1) & -83.2862(4) \\
	\hline\hline
    \end{tabular}
    \label{tab:WO_arep_gs}
\end{table}
For the $^3\Sigma$ state, the PBE nodal surface results in the lowest energy overall by 2~mHa.
Clearly, the $^3\Sigma$ state is correctly predicted to be the ground state, even in the absence of spin-orbit.
Now that the ground state has been determined, we can determine the dissociation energy $D_e$ at equilibrium.
We calculate the W atom and O atom within FNDMC, and the dissociation energy can be determined by $D_e = E(\textrm{WO}) - E(\textrm{W}) - E(\textrm{O})$.
Note that our dissociation energy does not include the zero-point motion.
We find the dissociation energy to be $D_e^{FNDMC} = 6.80(2)$~eV. 
To see how this compares to the experimental results, we first note that the experimental vibrational frequency is $\omega_e = 1065.6231(52)$~cm$^{-1}$ \cite{ram2001} and the 
experimental binding energy (including zero point motion) is $D_0^{expt} = 6.85(44)$~eV \cite{nist}. 
Correcting for the zero point motion, the experimental dissociation energy becomes $D_e^{expt} = 6.91(44)$~eV.
Without the spin-orbit correction, the FNDMC dissociation energy falls well within the experimental error bars. 
Regarding the excited state $^5\Pi$, previous CASSCF (complete active space 
self-consistent field) calculations predicted a minimum a $r_e = 2.11$~\AA \,and $D_e$ = 2.42~eV whereas CISD found $r_e = 1.99$~\AA \, and $D_e = 3.25$~eV \cite{nelin_bauschlicher}.
The more recent CMRCI finds an equilibrium bond length which is shifted toward the $^3\Sigma$ minimum at $r= 1.72$~\AA.
We calculate the $^5\Pi$ state at this geometry within FNDMC using PBE0 nodes and find a total energy of -83.2848(3)~Ha. 
Comparing with the results in Table \ref{tab:WO_arep_gs}, it is clear that FNDMC does not predict $r_e = 1.72$~\AA\, to be the minimum for the $^5\Pi$ state, and in our 
calculations it has a lower energy at $r_e = 1.67$~\AA.

\begin{figure}[t]
    \centering
    \caption{FPSODMC total energies of
    WO molecule as a function of spin time step $\tau_s$/effective spin mass $\mu_s$ for the lowest 
    states with $^3\Sigma$ and $^5\Pi$ symmetries. Note that the within the error bars the difference between the states remains very similar regardless of the time step.
}   
\includegraphics[width=0.75\textwidth]{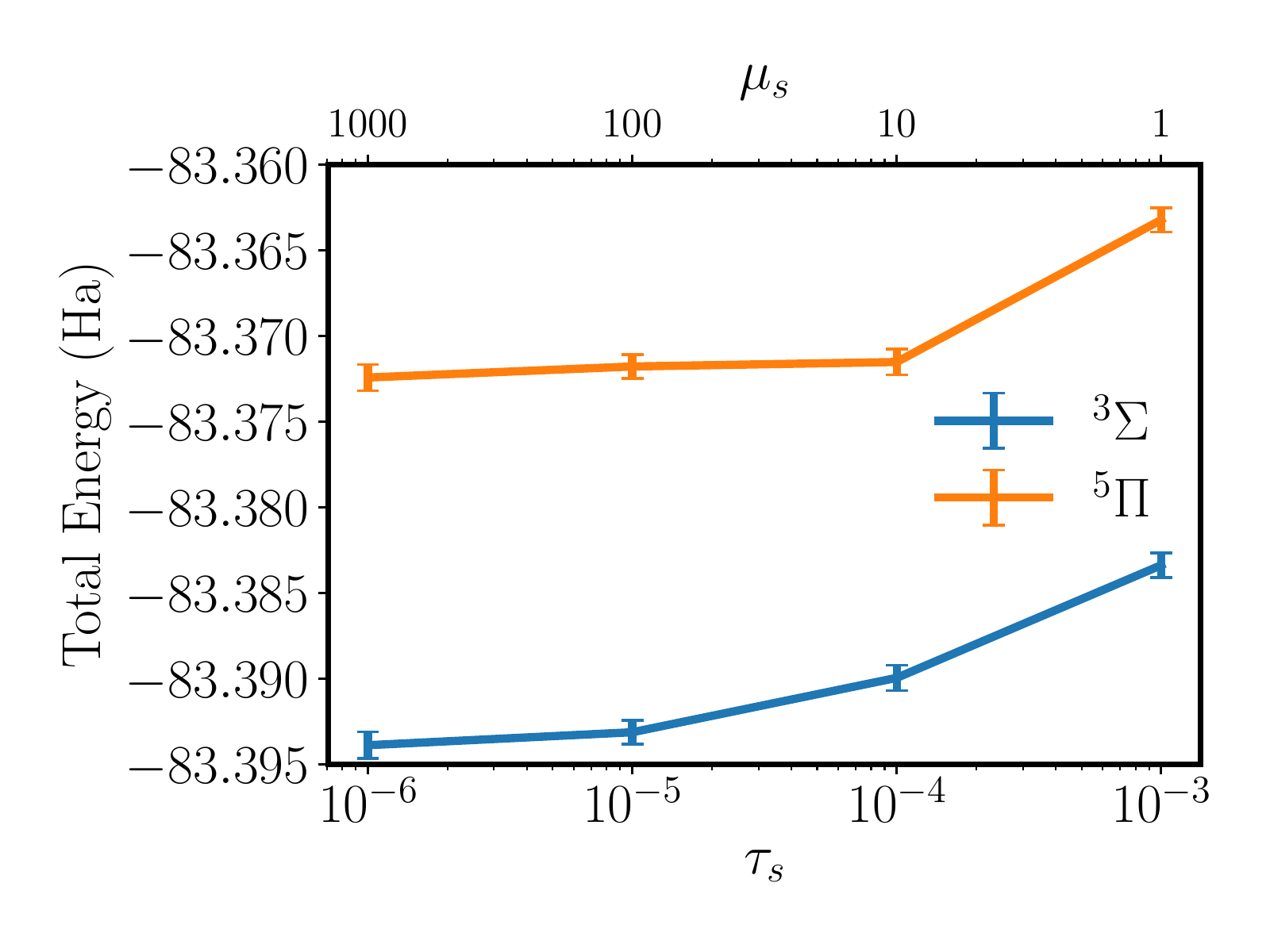}
    \label{fig:WO}
\end{figure}

We now turn to the FPSODMC results with explicit inclusion of spin-orbit and two-component spinors.  
With the spin-orbit interaction included, the $^3\Sigma$ and $^5\Pi$ states are split. 
We take our trial wave functions from the lowest energy spin-orbit state coming from each of the term symbols calculated from COSCI, which we indicate by $^3\Sigma$ and $^5\Pi$ respectively.
As above, we first investigate the difference between the ground states of the two manifolds at $r_e = 1.67$~\AA.
The results are shown in Figure \ref{fig:WO}.
As mentione above we extrapolate in the spin time step to find the desired expectation values. Note that the time step for the spatial degrees of freedom was conservatively small $\tau=10^{-3}$ a.u..
It is clear that 
 FPSODMC predicts the $^3\Sigma$ symmetry to be the ground state, in agreement with 
 CMRCI results quoted above.
In order to predict the dissociation energy, we calculate the ground states of individual atoms and find a dissociation energy of $D_e^{FPSODMC} = 6.52(2)$~eV. 
We note that the predicted dissociation energy for the case without spin-orbit and with spin-orbit differ by roughly $0.3$~eV.
While the quality of the nodal surface or phase can shift the predicted value slightly (we use DFT nodal surfaces whereas we use COSCI phases due to difficulties in generating an open-shell DFT trial wave function within the relativistic framework), another factor contributes more strongly to this difference. 
While the spin-orbit interaction tends to lower the ground state energy when it splits a term symbol into various states, this effect is much less pronounced in the WO molecular system. 
The spin-orbit splitting leads to lower 
energy for an individual atom, ultimately resulting in a reduced estimation of the dissociation energy than what the scalar relativistic FNDMC will predict.
Additionally, we note that the prediction from FPSODMC agrees with experiment to the experimental uncertainty.
We also investigate the location of the $^5\Pi$ minimum.
At $r_e = 1.72$~\AA\, as predicted by CMRCI, FPSODMC finds a total energy of -83.3705(5)~Ha. 
If we compare to the total energy shown in Figure \ref{fig:WO} which saturates to -83.3724(7)~Ha, we find that $r_e = 1.72$~\AA\,  is not the equilibrium geometry for this excited state, in agreement with the scalar relativistic FNDMC calculations.

\subsection{Tungsten Dimer}
Unlike WO which differs from its isovalent counterparts, W$_2$ is expected to follow the trend of Mo$_2$ and Cr$_2$ and form a $^1\Sigma$ ground state configuration.
To the best of our knowledge, experimental data does not exist for the binding energy of the W$_2$.
However a number of model estimates are in the literature predicting a wide range of binding energies in eV, namely 4.69(89) \cite{dimer1}, 5.62(1.23) \cite{dimer2}, 4.68 \cite{dimer3} and 5.00(69) \cite{dimer4}. 
From the scatter in the various estimates, an adopted prediction of the binding energy is given as $D_0^{est} = 5(1)$~eV \cite{morse}.
More recent B3LYP calculations predict a $^1\Sigma$ ground state over a $^3\Sigma$ calculation with an equilibrium bond length of 2.048~\AA\, \cite{kraus}.
Multi-reference perturbation theory estimates an equilibrium bond length of $r_e = 2.0561$~\AA\, and $D_e$ = 4.5110~eV \cite{angeli} in the $^1\Sigma$ state.
Furthermore, a CASPT2 study found a dissociation energy of 5.37~eV and $r_e = 2.010$~\AA\, \cite{borin}. 
We aim to add DMC among the predictions of the dissociation energy for W$_2$. 
A summary of the predictions are listed in Table \ref{tab:w2}.
Although there is consistent discrepancy among various theories for the dissociation energies, the vibrational frequencies are in reasonable agreement and are quite small.
Experimental detection of W$_2$ in an argon matrix found $\omega_e$ = 336.8~cm$^{-1}$ \cite{hu}, and theoretical estimates include $\omega_e = 401.2$~cm$^{-1}$ \cite{kraus}, $\omega_e = 326.69$~cm$^{-1}$ \cite{angeli}, and $\omega_e = 354$~cm$^{-1}$ \cite{borin}. 
Because of the very small vibrational frequencies, the binding well is relatively shallow and energetic changes for geometric differences on the order of $0.01$~\AA\, as is seen between the various other predictions, will be difficult to discern within statistical errors within DMC, so we adopt a geometry of the intermediate bond length of $r = 2.048$~\AA\, to perform all our calculations.

\begin{figure}[t]
    \centering
    \caption{The difference in the charge density between DHF and PBE0 trial wave functions for the outermost spinors $\pi^4 \sigma_{z^2}^2 \sigma_s^2 \delta^4$. In red, we show where PBE0 has a greater charge density and in blue where DHF shows a greater charge density. The black spheres indicate the location of the Tungsten atoms.  }
    \input{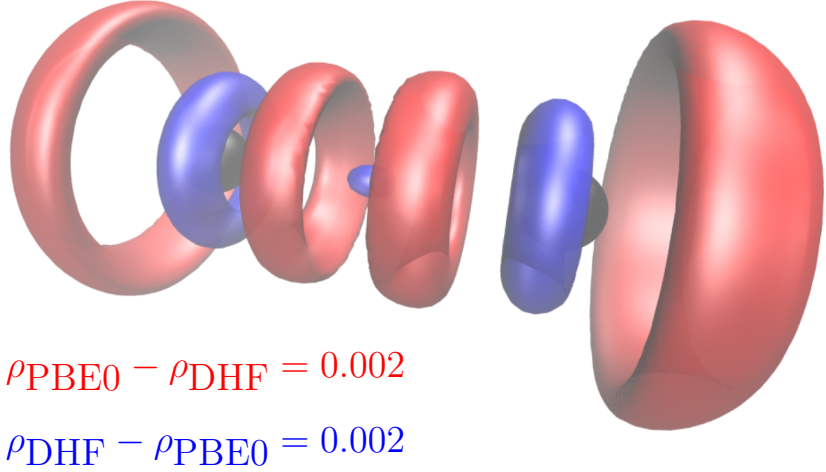}
    \label{fig:density}
\end{figure}

\begin{table}[!h]
   \centering
   \caption{FNDMC (AREP) and FPSODMC (REP)  dissociation energies of W$_2$ compared with other methods. In the fixed-node DMC calculations the trial nodes are from Slater determinants built from PBE0 single-particle orbitals. }
    \begin{tabular}{lccr}
       \hline\hline
       Method & $D_0/$eV   & $r_e$/\AA & Ref.    \\
       \hline
       FNDMC/PBE0$^\dagger$          &  5.34(2)   & 2.048$^*$  &  This Work \\
       PBE$^\dagger$            &  4.57      & 2.048$^*$  &  This Work \\
       PBE0$^\dagger$           &  2.54      & 2.048$^*$  &  This Work \\
       B3LYP$^\dagger$          &            & 2.048      & \cite{kraus} \\
       SC-NEVPT3$^\dagger$      &  4.5110    & 2.056      & \cite{angeli} \\
       CASPT2 $^\dagger$        &  5.37      & 2.010      & \cite{borin} \\
       FPSODMC/DF$^\ddagger$    &  2.51(2)   & 2.048$^*$  & This Work \\
       FPSODMC/COSCI$^\ddagger$ &  3.03(2)   & 2.048$^*$  & This Work \\
       FPSODMC/PBE0$^\ddagger$  &  4.17(2)   & 2.048$^*$  & This Work \\
       Estimated                &  5(1)      &            & \cite{morse} \\
       \hline
    \end{tabular}\\
$^*$ Indicates the calculation was performed at this geometry. \\
$^\dagger$ Indicates a spin-averaged calculation\\
$^\ddagger$ Indicates an explicit spinor calculation
    \label{tab:w2}
\end{table}

We now turn to our DMC results.
Utilizing a scalar relativistic Hamiltonian within FNDMC, we test both a PBE and PBE0 nodal surface in order to minimize the DMC energy.
PBE0 results in a lower energy by 3~mHa, and produces a dissociation energy of 5.34(2)~eV. 
Since W$_2$ is a closed shell molecule, when we include spin-orbit within FPSODMC we first try a single slater determinant built from DHF spinors.
However, this only results in underbinding with dissociation energy of 2.51(2)~eV that
results from rather poor approximation of the phase and consequently a larger fixed-phase error.  
We attempt to improve the fixed-phase approximation by using the COSCI expansion where the anti-bonding spinors are included in the active space.
While this lowers the total energy within FPSODMC, the dissociation energy increases only to 3.03(2)~eV. As has been seen in other transition metal systems \cite{wagner07}, HF orbitals tend to localize charge too much 
close to the ions 
\cite{wagner2007a,kolorenc2010}.
This is due to the fact that the only way for HF to lower the energy is through exchange, which becomes larger when the charge density
and particular states become more localized.
The same bias is then built-in also in the excited virtual one-particle space.
This results in a well-known
 "too ionic" bias of the HF or HF-based wave functions
 and it is expected that the same persists
 with spin-orbit included. 
We therefore consider further possibilities for improving the trial phases here.
Since W$_2$ is a closed-shell molecule, we can utilize a single Slater determinant built from DFT relativistic spinors. 
In Figure \ref{fig:density}, we illustrate the difference in the relativistic charge density between the PBE0 and DHF trial wave functions. 
As in the case of other FNDMC treatments of transition metals, HF localizes more on the atoms whereas the PBE0 pushes more charge density into the bonding region. 
For the PBE0 trial phase, we find a significantly lower energy resulting in a dissociation energy of 4.17(2)~eV. 
To ensure the quality of the trial phase, we also try LDA, PBE, and B3LYP trial wave functions. 
All of these result in the same total energy as PBE0 within statistical error bars. 
Note that compared to the FNDMC without spin-orbit, there is a significant contribution to the dissociation energy coming from spin-orbit.

\section{Conclusions}
\label{sec:conclusions}
We have presented our recent development for QMC method with Hamiltonians that include particle spins. We applied these developments to two molecules with tungsten, WO and W$_2$, and we compared the results obtained with traditional spin-orbit averaged FNDMC and 
with explicit treatment of spin-orbit in
the FPSODMC methods.
We illustrate that DMC is able to deal with systems containing heavy elements by using both scalar relativistic as well as spin-orbit interactions through the use of non-local ECPs and that the results enable to 
see the differences that result from 
proper treatment of spins as quantum variables.

For WO, we find that the ground state is in agreement with the experimental determination, and both the scalar relativistic and spin-orbit prediction for the dissociation energy are in agreement with the (large) experimental error. 
We also show that previous theoretical estimates for the excited state predict an equilibrium bond length appears larger by a few percent.  
For W$_2$, there very limited amount of experimental data to compare against.
Instead, we make a genuine prediction for the binding energy.
Additionally we show that for the heavy transition metals, trial phases that localize charge density too much on the atoms results in a larger fixed-phase bias. This agrees 
with previous investigations in transition 
metal systems where the use of the DFT/hybrid functionals generally improved the one-particle orbitals by easing too ionic
character of Hartree-Fock based orbitals \cite{wagner2007a, kolorenc2010}.

DMC has been one of the most accurate methods for treating strong correlations in various materials, and with the addition of spin-orbit it can now treat heavy-element compounds to high accuracy. 
We note however that in order to reach higher accuracies, various improvements need to be addressed. In particular, 
more accurate relativistic ECPs that are designed to be used in a correlated framework with much higher accuracy and better benchmarking will be needed.  
Additionally, higher accuracy trial wave functions will be important. This 
will presumably involve combinations of  configuration interaction expansions,
improved open-shell DFT trial wave 
functions and their combinations. Nevertheless, the presented developments 
open new possibilities for accurate many-body calculations of systems with heavy atoms 
and significant spin orbit-effect by 
accurate QMC methods.\\
\bigskip

\noindent
{\bf Acknowledgements. }
This research was supported by the U.S. Department of Energy (DOE), Office of Science, Basic Energy Sciences (BES) under Award de-sc0012314. For calculations we utilized Stampede at TACC.

\section*{References}
\bibliographystyle{model1-num-names}
\bibliography{refs}

\end{document}